\documentclass[letter]{aa} % for the letters 
\usepackage{upgreek}

\usepackage[varg]{txfonts}
\usepackage{hyperref}
\usepackage{graphicx}

\hypersetup{%
  colorlinks=true,% hyperlinks will be black
  linkcolor=blue,% hyperlink borders will be red
  citecolor=blue
}

\newcommand{\OI}{O\,{\sc i}}
\newcommand{\CI}{C\,{\sc i}}
\newcommand{\SII}{S\,{\sc ii}}

\newcommand{\CII}{C\,{\sc ii}}
\newcommand{\HII}{H\,{\sc ii}}
\newcommand{\HI}{H\,{\sc i}}

\begin{document}

\title{The initial  gas-phase sulfur abundance in  the Orion Molecular Cloud from sulfur radio recombination lines\thanks{Based on observations with the 40m radio telescope of the \mbox{Spanish National} Geographic Institute (IGN) at Yebes Observatory 
\mbox{(project 19A005)}.
Yebes Observatory thanks the ERC for funding \mbox{support} under grant ERC-2013-Syg-610256-NANOCOSMOS.}}

%%\subtitle{}

 \titlerunning{The sulfur abundance in the Orion Molecular Cloud} 
\authorrunning{Goicoechea \& Cuadrado} 
                                      
 \author{Javier\,R.\,Goicoechea\inst{1}
          \and
        Sara Cuadrado\inst{1}}

 \institute{Instituto de F\'{\i}sica Fundamental
     (CSIC). Calle Serrano 121-123, 28006, Madrid, Spain.
              \email{javier.r.goicoechea@csic.es}}
   
   \date{Received 8 February 2021 / Accepted 2 March 2021}

% \abstract{}{}{}{}{}     

% Context.
% Aims.
% Methods.
% Results.
% Conclusions.

\abstract{The abundances of chemical elements and their depletion factors  are
essential parameters  for understanding the composition of the gas and dust that are ultimately incorporated into stars and planets.
Sulfur is an abundant but peculiar element in the sense that, despite being less volatile than other elements (e.g., carbon), it is not a major constituent of dust grains in diffuse interstellar clouds.
Here, we  determine the gas-phase carbon-to-sulfur abundance ratio, \mbox{[C]\,/\,[S]}, and the  [S] in a dense star-forming cloud  from new radio recombination lines (RRLs) detected with the Yebes\, 40m telescope --  at relatively high frequencies \mbox{($\sim$\,40\,GHz\,$\simeq$\,7\,mm)} and angular resolutions (down to 36$''$) --  in the Orion Bar, a rim of the Orion Molecular Cloud (OMC). We detect nine C$n$$\alpha$ RRLs \mbox{(with \mbox{$n$\,=\,51 to 59})} as well as nine narrow line   features separated from the C$n$$\alpha$ lines by \mbox{$\delta$$v$\,=\,$-$8.4\,$\pm$\,0.3\,km\,s$^{-1}$}.  Based on this  velocity separation, we assign  these features to sulfur RRLs, with little contribution of  RRLs from the more condensable elements
\mbox{Mg, Si, or Fe}. Sulfur RRLs lines trace the   photodissociation region (PDR)
of the OMC. In these neutral gas layers, up to \mbox{$A_V$\,$\simeq$\,4}, the ions C$^+$ and S$^+$  lock in most of the  C and S  gas-phase reservoir. 
We determine a relative abundance of \mbox{[C]$_{\rm Ori}$\,/\,[S]$_{\rm Ori}$\,=\,10.4\,$\pm$\,0.6} and, 
adopting the same [C]$_{\rm Ori}$   measured  in the translucent  gas  toward  star  \mbox{$\theta^1$ Ori B}, an absolute abundance of [S]$_{\rm Ori}$\,=\,(1.4\,$\pm$\,0.4)$\cdot$10$^{-5}$.
This value is consistent with emission models of the
observed sulfur RRLs if \mbox{$N$(S$^+$)\,$\simeq$\,7$\cdot$10$^{17}$\,cm$^{-2}$} (beam-averaged). The
[S]$_{\rm Ori}$ is  the ``initial'' sulfur abundance in the OMC,
before an undetermined fraction of the [S]$_{\rm Ori}$ goes into  molecules and ice  mantles in the cloud interior. The inferred abundance [S]$_{\rm Ori}$ matches the solar abundance, thus  implying that there is little  depletion of sulfur onto rocky dust grains, with
\mbox{$D$(S)\,=\,0.0\,$\pm$\,0.2 dex}.
}


 \maketitle
%
%-------------------------------------------------------------------

%%%%%%%%%%%%%%%%%%%%%%
\section{Introduction}
%%%%%%%%%%%%%%%%%%%%%%  

 \mbox{Ultraviolet} (UV) line absorption studies show that the \mbox{gas-phase} abundance of  certain elements  (Si, Mg, Fe, Ca, Ti, etc.)
 are highly depleted  in  \mbox{diffuse}  (\mbox{$A_V$\,$\lesssim$\,1}) and 
\mbox{translucent (\mbox{$A_V$\,$\gtrsim$\,1})} low-density clouds of the 
    interstellar medium (ISM)\footnote{We refer to the gas-phase abundance ``[X]''  with respect to H nuclei   as the gas column density ratio \mbox{[X]\,=\,$N$(X)\,/\,$N_{\rm H}$}, where 
 $N_{\rm H}$\,=\,$N$(H)\,+\,2$N$(H$_2$) in  molecular clouds.
We denote the ``solar'' or ``bulk'' abundance of element X in the Sun (i.e., the current \mbox{photospheric} abundances corrected for \mbox{diffusion})  
as \mbox{[X]$_\odot$}, with 
\mbox{[C]$_\odot$\,=\,2.9$\cdot$10$^{-4}$}, 
\mbox{[Fe]$_\odot$\,=\,3.5$\cdot$10$^{-5}$}, 
\mbox{[Mg]$_\odot$\,=\,4.4$\cdot$10$^{-5}$}, and
\mbox{[Si]$_\odot$\,=\,3.6$\cdot$10$^{-5}$} \citep{Asplund09}.
We took these \mbox{[X]$_\odot$}  as the cosmic abundances and
used the \mbox{logarithmic} depletion factor defined as 
\mbox{$D$(X)\,=\,log\,[X]\,$-$\,log\,[X]$_\odot$}.}, with \mbox{$D$(X) ranging from $-$1 to $-$3\,dex} \citep[][]{Savage96,Sofia04b,Jenkins09}. This is consistent with their incorporation into (mainly) silicate  grains \citep[][]{Mathis90,Draine03}. 
The inferred depletion factors of  much more volatile elements, such as carbon,
are lower but still significant, \mbox{$D$(C)\,$\gtrsim$\,$-$0.5\,dex}.

\mbox{Sulfur} is on the top ten list of the most abundant  elements$^1$, with \mbox{[S]$_{\odot}$\,=\,(1.4\,$\pm$\,0.1)$\cdot$10$^{-5}$}, but it is  an unusual element in the sense that almost all of the \mbox{sulfur} in diffuse clouds 
remains in the gas phase \citep[][]{Fitzpatrick94,Howk06}. 
However,  the relevant UV \SII~lines  saturate, and thus 
the determination of  [S] in neutral atomic \HI~and translucent molecular clouds
may be uncertain \mbox{\citep[][]{Federman93,Sofia04b}}. 
  
Sulfur plays an important role in stellar nucleosynthesis; it is mainly produced in \mbox{massive} stars
\citep[e.g.,][]{Perdigon21},
in star formation, and in astrochemistry \citep[e.g.,][]{Fuente17,Shingledecker20}.  
Inside star-forming clouds, an undetermined \mbox{fraction} of sulfur  gradually converts into  
molecules and ice  mantles \mbox{\citep[e.g.,][]{Goico21}}, and yet the initial [S]
in these dense molecular  clouds is poorly constrained.  

\mbox{Radio} recombination lines (RRLs) 
provide a  powerful tool for studying star-forming regions
independently of dust obscuration. Hydrogen and He RRLs are extensively used to constrain the morphology and 
physical conditions ($T_{\rm e}$ and $n_{\rm e}$) of
fully \mbox{ionized} \mbox{\HII~regions} \mbox{\citep[e.g.,][]{Churchwell78}}.
However, only  stellar far-UV (FUV) photons, with energies below 13.6\,eV, permeate  the rims of  molecular clouds, so-called \mbox{photodissociation} regions \citep[PDRs;][]{Hollenbach_1999}. In the first layers of a PDR, the dominant state of  elements with ionization potential (IP) below H  is singly ionized: \mbox{C$^+$~(11.3\,eV)}, \mbox{S$^+$~(10.4\,eV)}, 
\mbox{Si$^+$~(8.2\,eV)}, \mbox{Fe$^+$~(7.9\,eV)}, or \mbox{Mg$^+$~(7.6\,eV)}. 
Indeed, the narrower carbon  RRLs  probe these neutral PDRs adjacent to  \mbox{\HII~regions} \citep[][]{Ball70,Natta94,Wyrowski97,Salas19,Cuadrado19}. 
Carbon RRLs have historically been detected in the centimeter (cm) \mbox{domain}: $\sim$8.6\,GHz for the C91$\alpha$ line 
(where $\alpha$ stands for \mbox{$\Delta$$n$\,=\,1} transitions).  
\mbox{Compared} to the fine-structure \mbox{$^2$$P_{3/2}$-$^2$$P_{1/2}$}  line of C$^+$, the
very important far-infrared  \mbox{[\CII]\,158\,$\upmu$m} cooling line   \citep{Hollenbach_1999}, carbon RRLs are optically thin and their intensity is proportional
to \mbox{$n_{\rm e}^{2}$\,$T_{\rm e}^{-1.5}$}. That is, they have different
excitation properties  than the \mbox{[\CII]\,158\,$\upmu$m} line 
\mbox{\citep[e.g., ][]{Natta94,Salas19}}.

The $^4S$ ground-electronic state of sulfur ions does not have low-lying fine-structure splittings. Hence, S$^+$  in  \mbox{neutral}
 gas cannot be detected\footnote{In  \HII~regions and IFs, S$^+$ can be detected  through the infrared \mbox{$^2$$P$--$^2$$D$}  lines at $\sim$1.03\,$\upmu$m \citep[][]{Walmsley00} and 
 in the visible through the  \mbox{$^2$$D$--$^4$$S$}   lines at  6718, 6733\,\AA~\citep[][]{Pellegrini09}.
 \mbox{The estimation} of [S] from these lines is sensitive
to $T_{\rm e}$ gradients, \mbox{extinction}, and ionization corrections 
\mbox{\citep[][]{Rudolph06}}. 
 The ground $^4$$S$  and excited  $^2$$D$ electronic states are separated by $\sim$\,21,400\,K. 
In PDRs and cold molecular clouds,  S$^+$  largely exists in the $^4$$S$ state.}
by \mbox{far-infrared} observations.
\mbox{Instead}, sulfur RRLs,  although much less studied, can be detected,
typically after long-time integrations, and probe the 
\mbox{``S$^+$ layer''} of a PDR\footnote{Carbon and sulfur RRLs were identified in the early 1970s. These detections implied the presence of what at that time was called ``\CII~and~\SII~regions''  in  \mbox{star-forming} clouds. The term ``PDR''  started to be used  in the late 1980s to model these 
\mbox{FUV-illuminated} regions.}. Early observations with cm-wave single-dish radio telescopes, at several-\mbox{arcminute} resolution,
 allowed the detection of emission features at the high-frequency shoulder of \mbox{carbon RRLs}. These features were associated with sulfur RRLs but can correspond
to a  superposition of RRLs from \mbox{low-IP} \mbox{elements} heavier than carbon 
\citep[][]{Chaission72,Chaisson74,Pankonin78,Falgarone78,Silverglate84,Vallee89,Smirnov95}.

\begin{figure}[t]
\centering   
\includegraphics[scale=0.44, angle=0]{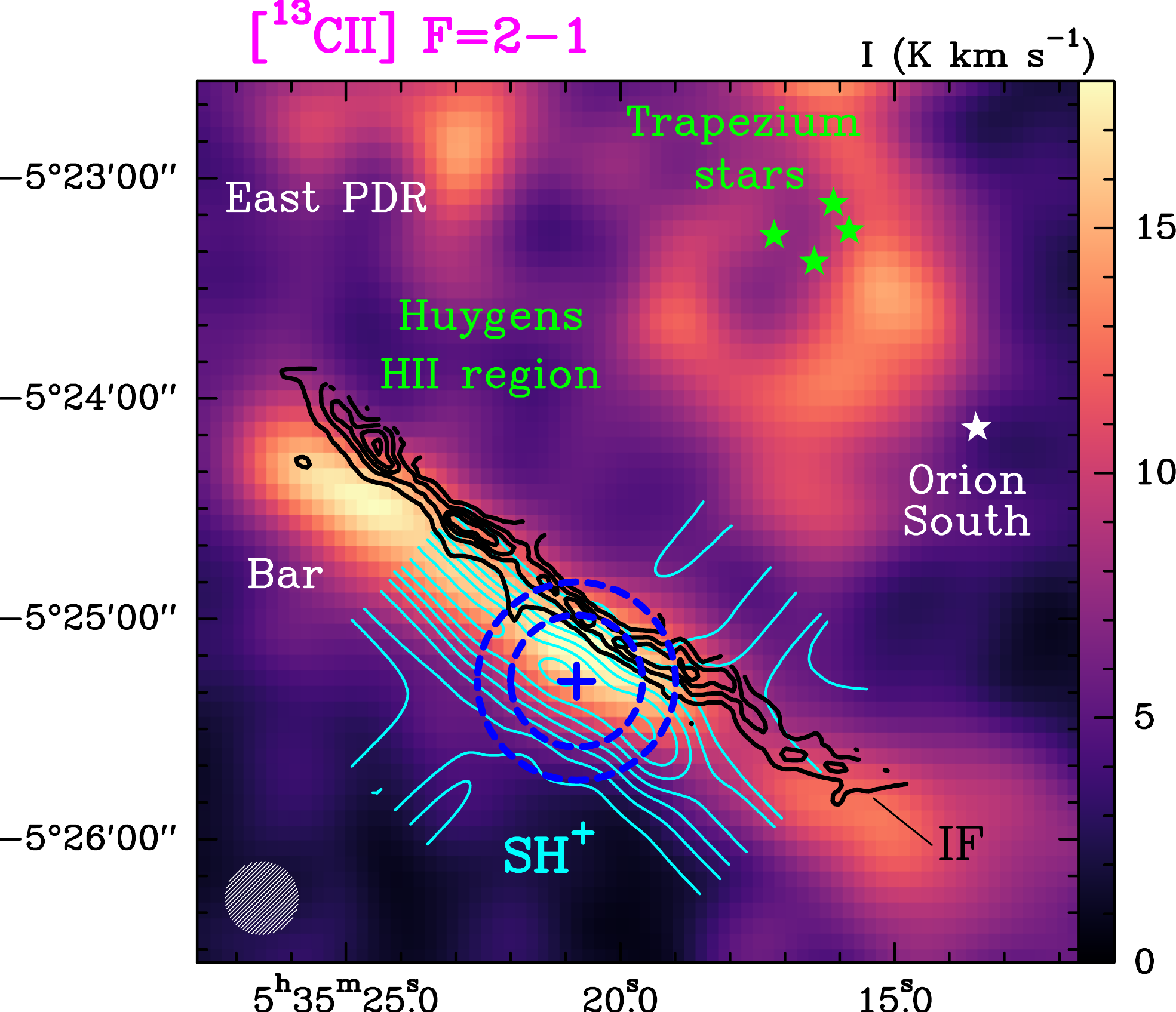}
\caption{Multiple \mbox{FUV-illuminated} edges in \mbox{OMC-1} (dense PDRs). The colored map shows the integrated emission ($I$) of the [$^{13}$\CII] \mbox{($^2$$P_{3/2}$-$^2$$P_{1/2}$  $F$\,=2-1)} line   at 1900.466\,GHz \citep[][]{Goico15}. This emission traces
the \mbox{C$^+$ layers} of the molecular cloud. Cyan contours 
show a  map of the \mbox{SH$^+$\,1$_0$-0$_1$ $F$\,=\,1/2-3/2}  emission  at 345.944\,GHz \mbox{\citep[][]{Goico21}}, a proxy of the \mbox{S$^+$ layer}
(both maps at $\theta_{\rm HPBW}$\,$\simeq$\,20$''$).
 Black contours show  the 
\mbox{\OI~\,1.317\,$\upmu$m fluorescent} emission  that delineates the 
the IF of the Bar \citep[][]{Walmsley00}. 
Dashed blue circles show the smallest and largest \mbox{$\theta_{\rm HPBW}$}
of our Yebes\,40\,m observations toward the DF position: 36$''$ and 54$''$,
respectively.} 
\label{fig:map}
\end{figure}

In this letter we present the first detection of sulfur RRLs in the Orion Bar 
\citep[e.g.,][]{Tielens93,Wyrowski97,Walmsley00,Goico16}. This \mbox{prototypical} dense PDR
  \mbox{(see Fig.~\ref{fig:map})}  is a
nearly edge-on  interface of the \mbox{Huygens}  \HII~region
 that is photoionized by the Trapezium massive stars \mbox{$\theta^1$ Ori}  \mbox{\citep[][]{Odell01}}  and the Orion Molecular Cloud (OMC) -- the closest region to host ongoing massive-star formation \mbox{\citep[e.g.,][]{Genzel89,Pabst20}}.

\vspace{-0.02cm}
\section{Observations}\label{sec:observations}

We used the Yebes\,40\,m radio telescope at Yebes Observatory (Guadalajara, Spain) to observe the  Bar. The observed position encompasses the dissociation front (DF), hereafter the \mbox{DF position,} 
at \mbox{$\mathrm{\alpha_{2000}=05^{h}\,35^{m}\,20.8^{s}\,}$}, \mbox{$\mathrm{\delta_{2000}=-\,05^{\circ}25'17.0''}$}.
We observed the complete $Q$\,band, from 31.1\,GHz to 50.4\,GHz, using the new high-electron-mobility transistor (HEMT) $Q$\,band receiver and fast Fourier transform spectrometers. These cover 18\,GHz of instantaneous bandwidth per polarization at a spectral resolution of 38\,kHz (\mbox{$\sim$\,0.3\,km\,s$^{-1}$} at \mbox{$\sim$\,40\,GHz}). We employed two frequency setups of slightly different central frequencies to identify any spurious line. The main beam efficiency varies from 0.6 at 32\,GHz to 0.43 at 48\,GHz, and the half power beam width (HPBW) at these frequencies ranges from \mbox{$\theta_{\rm FHWM}$$\sim$\,54$''$} to 36$''$  \citep{Tercero21}. 
We used the position switching observing mode, with a reference position  at an offset \mbox{($-$600$''$,\,0$''$)}.
The total integration time was $\sim$\,8\,h under winter conditions ($\sim$\,8\,mm of precipitable water vapor).
We reduced and analyzed the data  using the CLASS software of GILDAS. The achieved  root mean square (rms) noise  ranges from $\sim$\,3\,mK  to $\sim$\,10\,mK per velocity channel in antenna temperature units ($T^{*}_{\rm A}$).
Because of the relatively compact angular size  of the \mbox{C$^+$- and S$^+$-emitting} layers (compared to the secondary beam lobes), we converted the $T^{*}_{\rm A}$ scale to  main beam temperature, 
\mbox{$T_{\rm MB} = T^{*}_{\rm A}$$\cdot$ $\eta_{\rm F}/\eta_{\rm MB}$}, where $\eta_{\rm F}$ is the forward efficiency and $\eta_{\rm MB}$ the main beam efficiency. 
We adopted the values tabulated in \cite{Tercero21}: 
for example, $\eta_{\rm F}$\,$\simeq$\,0.92 and $\eta_{\rm MB}$\,$\simeq$\,0.52 at $\sim$\,40\,GHz. %%In addition to many other  lines that we will present elsewhere, 
We identified  nine C$n$$\alpha$ RRLs \mbox{(Table~\ref{Table_CRRL_Yebes})}  that show a fainter feature at slightly higher frequency, which we assigned to  S$n$$\alpha$ RRLs.
Figure~\ref{fig:all_RRLs} shows the individual spectra.

%-------------------------------------------------------------
%------------------------------------------------------------- 

\section{The C$^+$ and S$^+$ layers in the  Bar PDR}\label{sec-layers}

Figure~\ref{fig:map} shows the environment around the targeted position in the  Bar. The colored map shows the \mbox{$^2$$P_{3/2}$-$^2$$P_{1/2}$} fine-structure line emission (\mbox{$F$\,=2-1} hyperfine component) from the $^{13}$C$^+$ isotope, hereafter the 
\mbox{[$^{13}$\CII]\,158\,$\upmu$m line}.
While the \mbox{[$^{12}$\CII]\,158\,$\upmu$m} line is moderately optically thick
in dense PDRs \citep[e.g.,][]{Ossenkopf13}, \mbox{[$^{13}$\CII}] lines are
\mbox{optically} thin and  trace the  distribution of dense PDRs 
at the \mbox{FUV-illuminated rims of \mbox{OMC-1}} \citep{Goico15}.
 The ``\mbox{C$^+$ layer}'' (where most of the gas-phase carbon is in the form of C$^+$) extends from the \mbox{ionization front}  (IF; the   \mbox{H$^+$--H} transition at the \mbox{\HII~region--PDR} interface) to  the PDR layers, where most of the gas-phase \mbox{carbon} is converted first  into  C atoms and then into CO, at a few magnitudes of visual extinction  into the cloud ($A_V$). The distance between the IF and $A_V$(CO), perpendicular to the Bar and in the UV-illumination direction, is $\sim$15$''$  \citep[$\sim$0.03\,pc;][]{Tielens93,Goico16}. 
Interestingly, the distribution of the C91$\alpha$ emission, mapped %at $\sim$10$''$ resolution
 with the Very Large Array \citep[VLA;][]{Wyrowski97},  
is remarkably similar to the  \mbox{$\nu$\,=\,1-0 $S$(1)}  H$_2$ emission that traces the \mbox{H--H$_2$} transition at the DF.

Although S  has a lower IP than C, the extent of the \mbox{S$^+$ layer}  depends on  the wavelength- and $A_V$-dependent attenuation of the FUV field \citep{Goico07} 
as well as on  details of the S photoionization and   S$^+$ radiative recombination (RR)  and resonant \mbox{dielectronic} recombination (DR) processes. Paradoxically, a modern treatment of S$^+$ recombination, especially at low $T_{\rm e}$,   is still  lacking
\citep[][]{Badnell91,Bryans09}.
 Our observations encompass the \mbox{C$^+$} and \mbox{S$^+$ layers} of the Bar. Indeed, Fig.~\ref{fig:map}
shows that the observed DF position coincides with the position of the SH$^+$   emission peak (cyan contours). This trace molecular ion forms via gas reactions of S$^+$  with   vibrationally excited H$_2$ molecules \citep[e.g.,][]{Goico21}.

%-------------------------------------------------------------
\begin{figure}[t]
\centering   
\includegraphics[scale=0.5, angle=0]{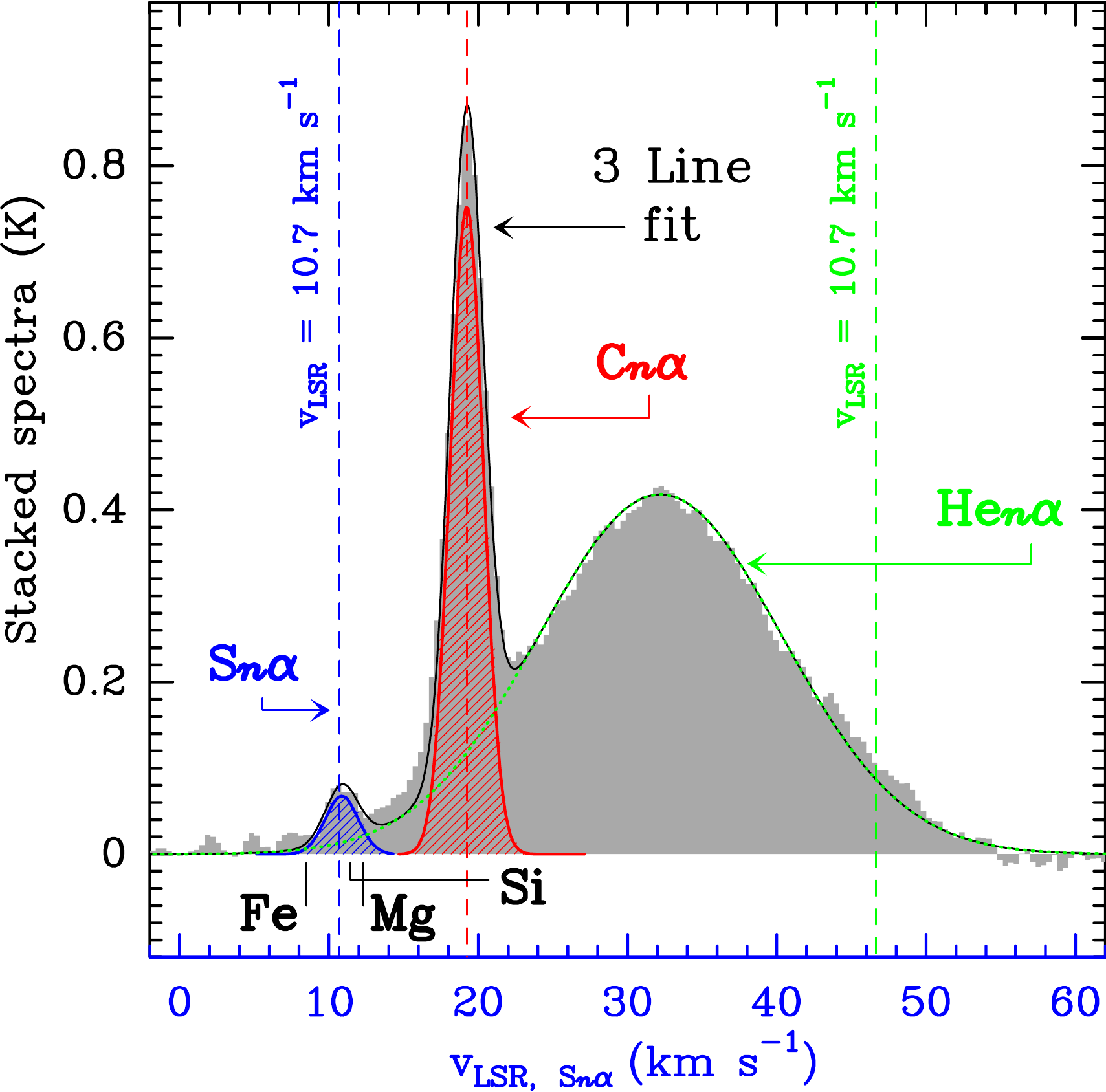}
%\vspace{-0.2cm}
\caption{Stacked spectrum \mbox{(filled gray histogram)} of several $n$$\alpha$ RRLs \mbox{($n$\,=\,54 to 59)}. The spectrum is centered at the frequencies of S$n$$\alpha$ lines and shows the expected position of Mg, Si, and Fe $n$$\alpha$ RRLs. The black curve is a three-Gaussian line fit: S$n$$\alpha$ (blue),
C$n$$\alpha$ (red), and He$n$$\alpha$ (green). 
Vertical dashed lines mark the LSR velocity of the Bar PDR.}
%\textcolor{red}{\mbox{i.e., He RRLs} are blueshifted with respect to the PDR emission}.}
\label{fig:stacked_spectra}
\end{figure}
%-------------------------------------------------------------

\section{Results: Detection of S$n$$\alpha$ RRLs}\label{sec-results}

We detected new C$n$$\alpha$ and He$n$$\alpha$ RRLs  (from \mbox{$n$\,=\,51 to 59};  \mbox{Fig.~\ref{fig:all_RRLs}}). Helium RRLs  
arise from the adjacent Huygens \mbox{\HII~region} \mbox{(IP$_{\rm He}$\,=\,24.6\,eV)} and from the different foreground layers of ionized gas in Orion's Veil \mbox{\citep{Odell01}}.The high electron temperature and \mbox{pressure} in the hot and fully ionized gas produce the broad line width  ($\Delta$$v$\,$\simeq$\,20\,km\,s$^{-1}$) of \mbox{He RRLs}. Consistent with ionized gas that flows toward the observer, \mbox{He RRLs} are blueshifted by  \mbox{$\sim$\,15\,km\,s$^{-1}$} with respect to OMC emission.

A visual inspection of each individual C$n$$\alpha$ spectrum shows the presence of a narrow emission feature at a slightly higher frequency. Unless produced by 
a molecular line\footnote{None of the S$n$$\alpha$ line frequencies correspond to molecules  present in the 
Bar \citep[][]{Cuadrado15,Cuadrado17} or in molecular-line catalogs.}, these features imply the presence of an RRL from an element heavier than 
\mbox{$^{12}$C (12\,amu)}. To determine their origin, we  
 calculated the $n$$\alpha$ RRL  frequencies of \mbox{$^{32}$S (31.972}\,amu) using the \mbox{Rydberg} equation  and assuming that the recombined electron in level $n$ moves on the field of a \mbox{nucleus} of effective charge \mbox{$Z_{\rm eff}$\,$\simeq$\,1\,=\,16\,($p^+$)$-$15\,(inner\,$e^-$)}. This  means that RRLs of \mbox{(many-electron)} neutral atoms
 possess frequencies close to hydrogen RRLs  but, as the mass of the nucleus increases, shifted to slightly higher frequencies 
 \mbox{(Table~\ref{Table_CRRL_Yebes})}. Likewise, we  computed the RRL frequencies of other neutral atoms with low IP ($<$13.6\,eV) and with  solar  abundances
[X]$_{\odot}$\,$>$\,10$^{-5}$: \mbox{$^{24}$Mg (23.985\,amu)}, 
\mbox{$^{28}$Si (27.977\,amu)}, and \mbox{$^{56}$Fe (55.935\,amu)}. 

In order to increase the signal-to-noise ratio (S/N) of the detection, and because RRLs of similar $n$ and $\Delta$$n$ have similar emission properties, we stacked all  C$n$$\alpha$  spectra from $n$\,=\,54 to 59 (those with the highest S/N).
We first converted each spectrum  
to the  local standard of rest (LSR) velocity scale,  then
resampled them to the same velocity channel resolution, and finally stacked them.
The resulting spectrum in  \mbox{Fig.~\ref{fig:stacked_spectra}} displays three line features that can be fitted  with three Gaussians reasonably well:  a broad line from He$n$$\alpha$ RRLs and two narrow lines, one from C$n$$\alpha$ RRLs and another from the heavier element.
According to the atomic mass differences between carbon and S, Mg, Si, and Fe, the velocity separation of their $n$$\alpha$ RRLs from  C$n$$\alpha$ lines is:  \mbox{$\delta$$v$($^{12}$C$-$$^{24}$Mg)\,=\,$-$6.9\,km\,s$^{-1}$},
\mbox{$\delta$$v$($^{12}$C$-$$^{28}$Si)\,=\,$-$7.8\,km\,s$^{-1}$}, \mbox{$\delta$$v$($^{12}$C$-$$^{32}$S)\,=\,$-$8.6\,km\,s$^{-1}$}, and \mbox{$\delta$$v$($^{12}$C$-$$^{56}$Fe)\,=\,$-$10.8\,km\,s$^{-1}$}.
The  observed \mbox{(peak-to-peak)} velocity separation  between the 
C$n$$\alpha$ and the  higher-frequency feature in the Bar  is \mbox{$\delta$$v$\,=\,$-$8.4\,$\pm$\,0.3\,km\,s$^{-1}$}. Therefore,
  the third  feature is largely produced by 
 \mbox{S$n$$\alpha$ RRLs}. This conclusion is supported by the severe depletions of  Mg, Si, and Fe  inferred in diffuse \mbox{neutral} clouds 
 \mbox{\citep[e.g.,][]{Jenkins09}}.
The typical depletion$^1$ factors in these clouds are \mbox{$D$(Mg)\,$\simeq$\,$-$1.3}, 
\mbox{$D$(Si)\,$\simeq$\,$-$1.4}, and \mbox{$D$(Fe)\,$\simeq$\,$-$2.2\,dex} 
\mbox{\citep[][]{Savage96}}. This implies 
\mbox{[S]\,$>$\,[Mg]\,$\simeq$\,[Si]\,$>$\,[Fe]} (in the gas phase). It also means that
in interstellar  PDRs,  RRLs of these more condensable elements  are much fainter than those of S. 
%if these \mbox{depletions} prevail in the Bar. 
From our observations we can only conclude that Mg, Si, and Fe are depleted by more than \mbox{$-$0.6 dex} relative to S.
 
The stacked S$n$$\alpha$ line shows a similar Gaussian profile,   
line width  (\mbox{$\Delta$$v$\,$\simeq$\,2.6\,$\pm$\,0.2\,km\,s$^{-1}$}; mostly microturbulent \mbox{broadening}), and 
velocity centroid 
(\mbox{$v$$_{\rm LSR}$\,$\simeq$\,10.8\,$\pm$\,0.1\,km\,s$^{-1}$})
as those of C$54$$\alpha$, [$^{13}$\CII], and \mbox{SH$^+$}  lines  -- all observed toward the DF position  at \mbox{$\theta_{\rm HPBW}$\,$\simeq$\,45$''$} (see \mbox{Fig.~\ref{fig:profiles}}). The narrow line profiles (see \mbox{Table~\ref{Table-fits}}) demonstrate that  S$n$$\alpha$ RRLs arise from PDR gas and not from the fully ionized \HII~region.

\section{Discussion}\label{sec-discussion}

\begin{figure}[t]

\centering   
\includegraphics[scale=0.43, angle=0]{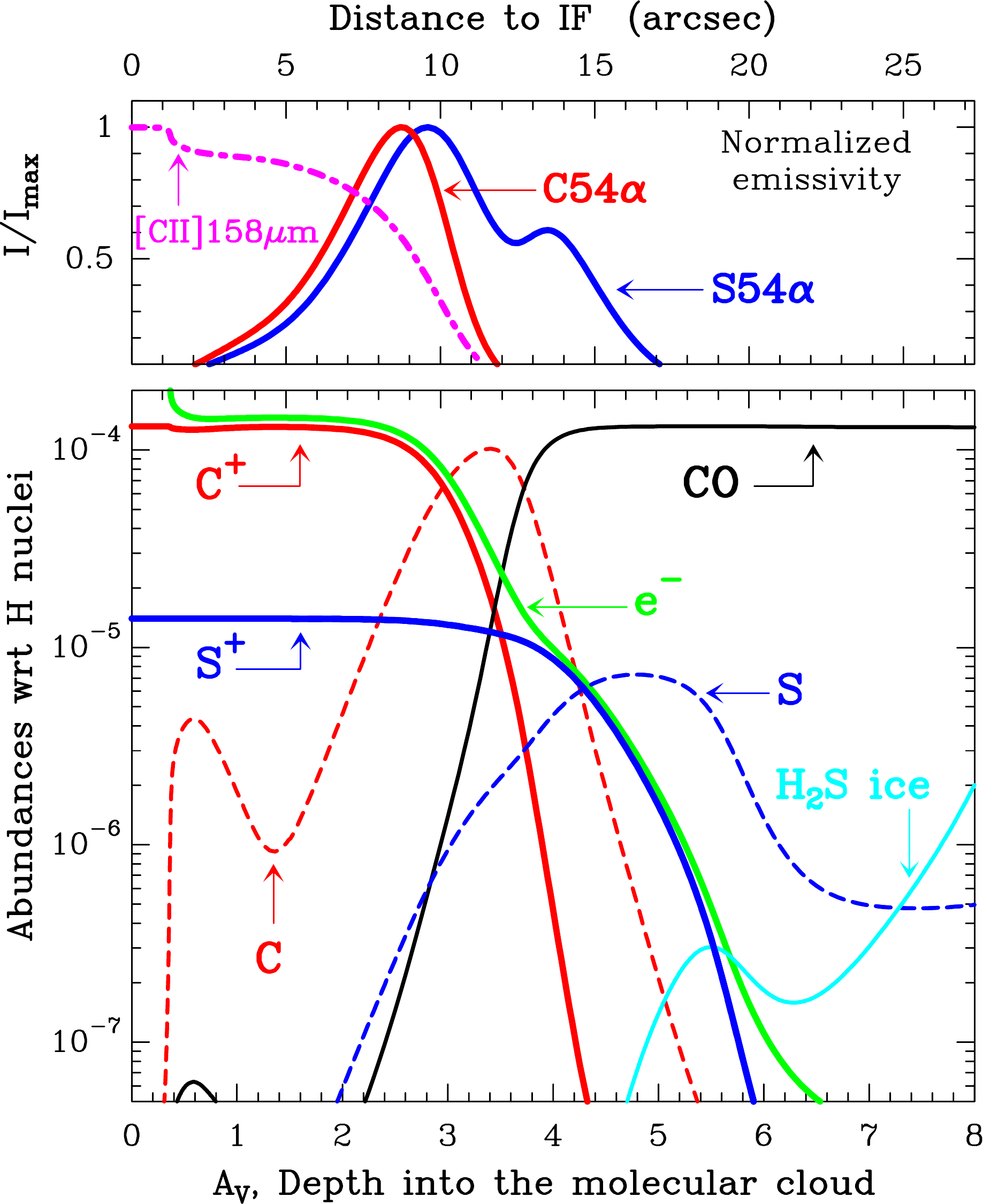}
\caption{Model$^5$ of the \mbox{C$^+$ layer} with
\mbox{[C]\,/\,[S]\,=\,10}. \textit{Lower panel}: Abundance profiles  as a function of $A_V$ along the \mbox{UV-illumination} direction:  from the IF to the  cloud interior  (lower axis) and  the equivalent angular distance  perpendicular to the Bar (upper axis). \textit{Upper panel:} Normalized  emissivities of the [\CII]\,158\,$\upmu$m, S54$\alpha$, and C54$\alpha$ lines.
For the RRLs, we  adopted an LTE  model (see Appendix~\ref{App-LTE-RRL}).}
\label{fig:pdr-model}
\end{figure}

To guide our interpretation,  Fig.~\ref{fig:pdr-model} shows the [C$^+$], [C], [CO], [S$^+$], [S], \mbox{[H$_2$S-ice]}, and [$e^-$]  abundance profiles  predicted
by a homogeneous PDR model\footnote{We adopted a uniform
gas  density \mbox{$n_{\rm H}$\,=\,$n$(H)\,+\,2$n$(H$_2$)\,=\,5$\cdot$10$^4$\,cm$^{-3}$}, an \mbox{FUV} radiation field   of \mbox{$\chi_{\rm FUV}$\,=\,10$^4$} times the Draine's field, and \mbox{$R_V$\,=\,$A_V$\,/$E_{B-V}$\,=\,5.5,} 
appropriate for Orion. 
This is sometimes called  the \mbox{``interclump''} medium of the Bar \mbox{\citep[e.g.,][]{YoungOwl00}}. This model is valid for the outer \mbox{C$^+$ layer}  and reproduces the observed \mbox{IF--$A_V$(CO)} angular separation.
The chemical network includes updated gas-phase and grain-surface reactions \citep[][]{Goico21}.}
 run with the Meudon code
\mbox{\citep{LePetit06}}.
We adopted the gas-phase carbon abundance, \mbox{[C]$_{\rm Ori}$\,=\,(1.4\,$\pm$\,0.6)$\cdot$10$^{-4}$}, measured 
 from UV absorption-line observations of the translucent neutral gas  toward the Trapezium star \mbox{$\theta^1$ Ori B} \mbox{\citep[][]{Sofia04}}. This abundance implies a depletion factor$^1$ of \mbox{$D$(C)\,$\simeq$\,$-$0.3\,dex}
 \citep[carbon that is \mbox{incorporated} into carbonaceous grains, 
 polycyclic aromatic hydrocarbons, and macromolecules such as
fullerenes; e.g.,][]{Berne12}.
 We assumed that this is precisely the initial gas-phase carbon abundance in the Bar
  \mbox{(i.e., [C$^+$]$_{\rm IF}$\,=\,[C]$_{\rm Ori}$)}.
Figure~\ref{fig:pdr-model} shows
that the C$^+$ abundance remains constant in the \mbox{C$^+$ layer}, from the IF
 (\mbox{$A_V$\,=\,0}) to \mbox{$A_V$(\CI)\,$\simeq$\,3}. 
The \mbox{S$^+$ layer} (where most of the sulfur is in the form of S$^+$) extends a bit deeper,  up to \mbox{$A_V$\,$\simeq$\,4}.
However, one would need observations at high angular resolution 
($\lesssim$\,5$''$) to spatially resolve the different sizes of the  \mbox{C$^+$} and \mbox{S$^+$ layers}; they are small  at the
 moderate gas densities of the Bar.
 In  these  intermediate PDR layers, \mbox{$A_V$\,$\simeq$\,3.5 to 5.5}, \mbox{photoionization} of S atoms becomes
the major source of $e^-$, and thus of the ionization fraction. This is one of the  reasons  why determining [S] in dense  gas is relevant.

  %-------------------------------------------------------------
%-------------------------------------------------------------

Because the  HPBW of the 40\,m telescope encompasses  the \mbox{C$^+$ and S$^+$ layers} of the Bar, and assuming that the C and S level populations of 
fairly high principal quantum numbers $n$  depart from local \mbox{thermodynamic} \mbox{equilibrium} (LTE) in a similar \mbox{fashion} \citep[e.g.,][]{Dupree74, Salgado17}, we can determine the  gas-phase \mbox{carbon-to-sulfur} abundance \mbox{ratio}, \mbox{[C]\,/\,[S]\,=\,$n_{\rm C^+}$/\,$n_{\rm S^+}$}, \mbox{directly} from the \mbox{C$n$$\alpha$\,/\,S$n$$\alpha$\,=\,$R_{n\alpha}$}
line intensity ratio. 
 From the stacked spectrum we obtain 
 \mbox{$R_{n\alpha}$\,=\,10.4\,$\pm$\,0.6}. This is a factor of two lower than the 
 \mbox{[C]$_{\odot}$\,/\,[S]$_{\odot}$ ratio}.
  As we adopted
\mbox{[C$^+$]$_{\rm IF}$\,=\,[C]$_{\rm Ori}$}, the  $R_{n\alpha}$ ratio
 leads to  
 \mbox{[S$^+$]$_{\rm IF}$\,=\,[S]$_{\rm Ori}$\,=\,(1.4\,$\pm$\,0.4)$\cdot$10$^{-5}$}.
 
Under the above assumptions, the initial [S]  in the OMC 
 matches the solar abundance. Interestingly, it also coincides with the stellar [S] abundances measured in young B-type stars of the \mbox{Ori OB1} association \citep{Daflon09}.
 This result implies
very little depletion \mbox{of S} in  dense clouds, \mbox{$D$(S)\,=\,0.0\,$\pm$\,0.2}.  In the unlikely case that the [C$^+$]$_{\rm IF}$ in the Bar  is significantly lower than the value measured by \cite{Sofia04},  our  estimated [S]$_{\rm Ori}$ 
 abundance  would be an upper limit and $D$(S) would decrease accordingly.
Still, an LTE excitation model 
with \mbox{$N$(S$^+$)\,$\simeq$\,7$\cdot$10$^{17}$\,cm$^{-2}$}
(\mbox{Appendix~\ref{App-LTE-RRL}}) is consistent with the absolute intensities 
 of the observed  sulfur RRLs, with \mbox{[S]$_{\rm Ori}$\,=\,1.4$\cdot$10$^{-5}$}, and with  \mbox{$N$(C$^+$)\,/\,$N$(S$^+$)\,$\simeq$\,10}.

 The  ratio $R_{\rm (54-59)\alpha}$ observed in the Bar 
 is higher than that determined at  several-arcminute resolution in the pioneering cm-wave observations  of the $\rho$\,Oph dark cloud  \citep[$R_{\rm 158\alpha}$\,$\simeq$\,3.5 and  $R_{\rm 110\alpha}$\,$\simeq$\,6.2;][]{Pankonin78} and 
 the W3A and Orion~B clouds
\citep[$R_{\rm 158\alpha}$\,$\simeq$\,3.1 and \,7.1, respectively;][]{Chaission72}.
These variations can be produced by \textit{i)} the different beam filling factors
of the  \mbox{C$^+$- and S$^+$-emitting} regions  in these clouds, such that their RRL emission is not spatially resolved and the large beam  mixes various PDRs, 
\textit{ii)} very different excitation properties of S$n$$\alpha$ versus C$n$$\alpha$ RRLs, or \textit{iii)} intrinsically  lower \mbox{[C]\,/\,[S]} ratios.
We currently favor possibility \textit{i)}. Indeed, 
 our  higher angular resolution observations  nearly spatially resolve the Bar PDR. 
If this were not the case, and if the \mbox{S$^+$ layer}  were significantly more extended than the \mbox{C$^+$ layer}, then the intrinsic $R_{\rm (54-59)\alpha}$ ratios would be
 larger. % (the beam filling correction for the  C$n$$\alpha$ lines would be larger).
In the future, more sensitive observations and more details on the  RR and DR of S$^+$ ions will allow us to  model the \mbox{non-LTE} excitation of sulfur RRLs individually, as well as to constrain $n_{\rm e}$ and $T_{\rm e}$ as a function of $A_V$  (e.g., \mbox{RRLs} do not arise from the same [\CII]\,158\,$\upmu$m-emitting  layer; see Fig.~\ref{fig:pdr-model}). In the meantime, ours is probably the most direct  estimation of \mbox{[S]} in a dense cloud.
 
 Since stellar sulfur abundances increase with increasing metallicity \citep[e.g.,][]{Perdigon21}, and the estimated [S] in stars and \mbox{\HII\,\,regions}  decreases  with galactocentric radius ($R_{\rm GC}$)
as $-$0.04\,dex\,kpc$^{-1}$ 
 \citep[e.g.,][]{Rudolph06,Daflon09,Arellano20}, we \mbox{anticipate} shallow variations of [S] in molecular clouds of different $R_{\rm GC}$.
  A related open question \citep[e.g.,][]{Sofia04b} is how to reconcile the
low sulfur depletion  in the ISM
with the existence of interplanetary dust particles containing sulfur \mbox{\citep[][]{Bradley94}} and with the existence of solid MgS
 in evolved stars. The latter seems only viable  if MgS is 
 present   in the outer coating surfaces of circumstellar grains and not in their cores \citep[][]{Lombaert12,Sloan14}. These  outer surfaces may be more easily destroyed in the harsh ISM.
 Observations of S$n$$\alpha$ RRLs from a larger sample of star-forming regions
 will improve our understanding of interstellar sulfur.

\begin{acknowledgements}  
We thank J. H. Black for reminding us that our spectra of the Orion Bar
may contain  sulfur RRLs and for useful remarks
on how S$^+$ \mbox{dielectronic} capture may proceed. We thank B. Tercero  and the Yebes
staff for their help with these observations.
 We thank the Spanish MCIYU for funding support under grants
\mbox{PID2019-106110GB-I00} and \mbox{AYA2016-75066-C2-1-P}.

\end{acknowledgements}

% WARNING
%-------------------------------------------------------------------
% Please note that we have included the references to the file aa.dem in
% order to compile it, but we ask you to:
%
% - use BibTeX with the regular commands:
%   \bibliographystyle{aa} % style aa.bst
%   \bibliography{Yourfile} % your references Yourfile.bib
%
% - join the .bib files when you upload your source files
%-------------------------------------------------------------------

\bibliographystyle{aa}
\bibliography{references}

\begin{appendix}\label{Sect:Appendix}

\section{Absolute intensities of sulfur RRLs}\label{App-LTE-RRL}  

The level populations of the recombined S atom 
can be written as \mbox{$N$($n$)\,=\,$b_n$\,$N$($n$)$_{\rm LTE}$},
where  $n$ is the principal quantum number, $N$($n$)$_{\rm LTE}$ are the LTE populations given by the \mbox{Saha-Boltzmann} equation,
and $b_n$ are the so-called departure coefficients. We consider  that the \mbox{S54$\alpha$ RRL} emission is dominated by 
 spontaneous emission, with little  contribution from stimulated emission.
Hence, we neglected amplification of any background \mbox{continuum} emission 
($T_{\rm c}$). This is a reasonable assumption for these RRLs in Orion
\citep[e.g.,][]{Natta94}.
Indeed, the 7\,mm continuum emission is produced by the
cosmic microwave \mbox{background}, the long-wave tail of the dust thermal emission,
and the \mbox{free-free} Bremsstrahlung emission. In Orion, the last component largely arises from the Huygens \HII~region, but this \mbox{nebular} emission is in the foreground relative to the Bar. 
In addition, owing to the low electron densities in PDRs, we estimate that the  free-free opacity  is very low, \mbox{$\tau_{\rm ff}$(7\,mm)\,$\sim$\,10$^{-7}$}.  Hence, the continuum opacity 
\mbox{($\tau_{\rm c}$\,=\,$\tau_{\rm ff}$\,$+$\,$\tau_{\rm d}$)}
 is dominated by the opacity of warm ($T_{\rm d}$\,$\simeq$\,50\,K) dust grains, which is also low,  $\tau_{\rm d}$(7\,mm)\,$\sim$\,10$^{-5}$. The observed sulfur \mbox{RRLs} are optically thin ($\tau_{\rm RRL}$ of  a few 10$^{-4}$) and are in the Rayleigh–Jeans 
regime (with $h\nu/k\ll T_{\rm e}$ and $T_{\rm c}$\,$\ll$\,$T_{\rm e}$). Hence, we
can   write the  S54$\alpha$ \mbox{integrated} line intensity  ($I_{\rm S54\alpha}$ in \mbox{K\,km\,s$^{-1}$}) as:
\begin{equation}\label{eq-I}
I_{\rm S54\alpha}=\displaystyle{\int} T_{\rm MB}\,{\rm {d}} v \simeq 15\,b_{54+1}\,T_{\rm e}^{-1.5}\,{\rm {exp\,(\chi_{54}})}\,EM_{\rm {S^+}},
\end{equation}
where \mbox{$\chi_n$\,=15.79$\cdot$10$^4$\,$n^{-2}$\,$T_{\rm e}^{-1}$} and \mbox{$EM_{\rm {S^+}}$\,=\,$n_{\rm e}$\,$n_{\rm S^+}$\,$L$} (in pc\,cm$^{-6}$)
is the S$^+$ emission measure along a slab of thickness $L$ 
(along the line-of-sight; $L$ should not be confused with the cloud depth in the \mbox{UV-illumination} direction). 
The factor $\simeq$\,15 is specific to 54$\alpha$ RRLs and includes the oscillator strength \mbox{$\Delta$$n$$M(\Delta$$n$)\,$\simeq$\,0.1908 for $\Delta$$n$\,=\,1}
\citep[][]{Brocklehurst71,Dupree74,Salas19}. When excitation conditions are close to LTE and as the principal quantum number $n$ increases, the departure coefficients 
approach \mbox{$b_n$\,$\simeq$\,1}. Coefficients for hydrogen and carbon RRLs have been estimated in the literature.
At the relatively low electron  temperatures of a PDR, these coefficients are typically \mbox{0.3\,$<$\,$b_n$\,$<$\,1}, with values  
up to $b_n$\,$\simeq$1.5 when the details of \mbox{resonant} dielectronic capture for \mbox{non-hydrogenic} atoms are included \mbox{\citep[e.g.,][]{Salgado17}}.
\mbox{In this study} we implicitly assumed that any substantial difference
between S$^+$ and C$^+$ \mbox{recombination} will have  little effect on the
relative intensities of their moderately high-$n$ recombination spectra.
Nonetheless, we point out again that the low $T_{\rm e}$ RR and DR of S$^+$ have not been properly studied  \citep[][]{Bryans09}.
Here, we simply adopted 
 \mbox{$b_n$\,=\,1} and \mbox{$T_{\rm e}$\,=\,$T_{\rm gas}$} (i.e., LTE)
to estimate the  RRL emissivities (see the PDR model in Fig.~\ref{fig:pdr-model}).

\begin{figure}[t]
\centering   
\includegraphics[scale=0.39, angle=0]{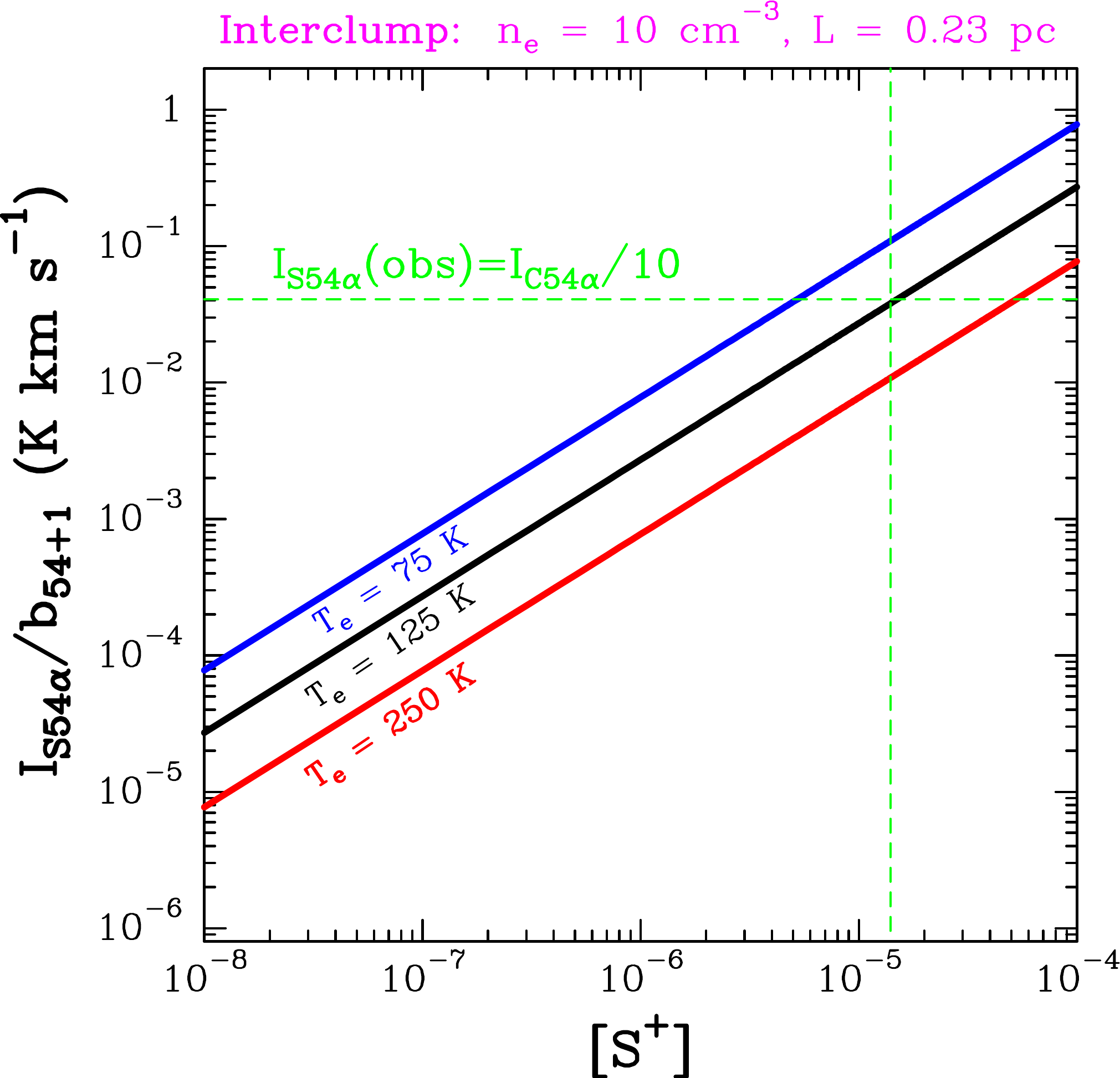}

\vspace{0.1cm}
\includegraphics[scale=0.39, angle=0]{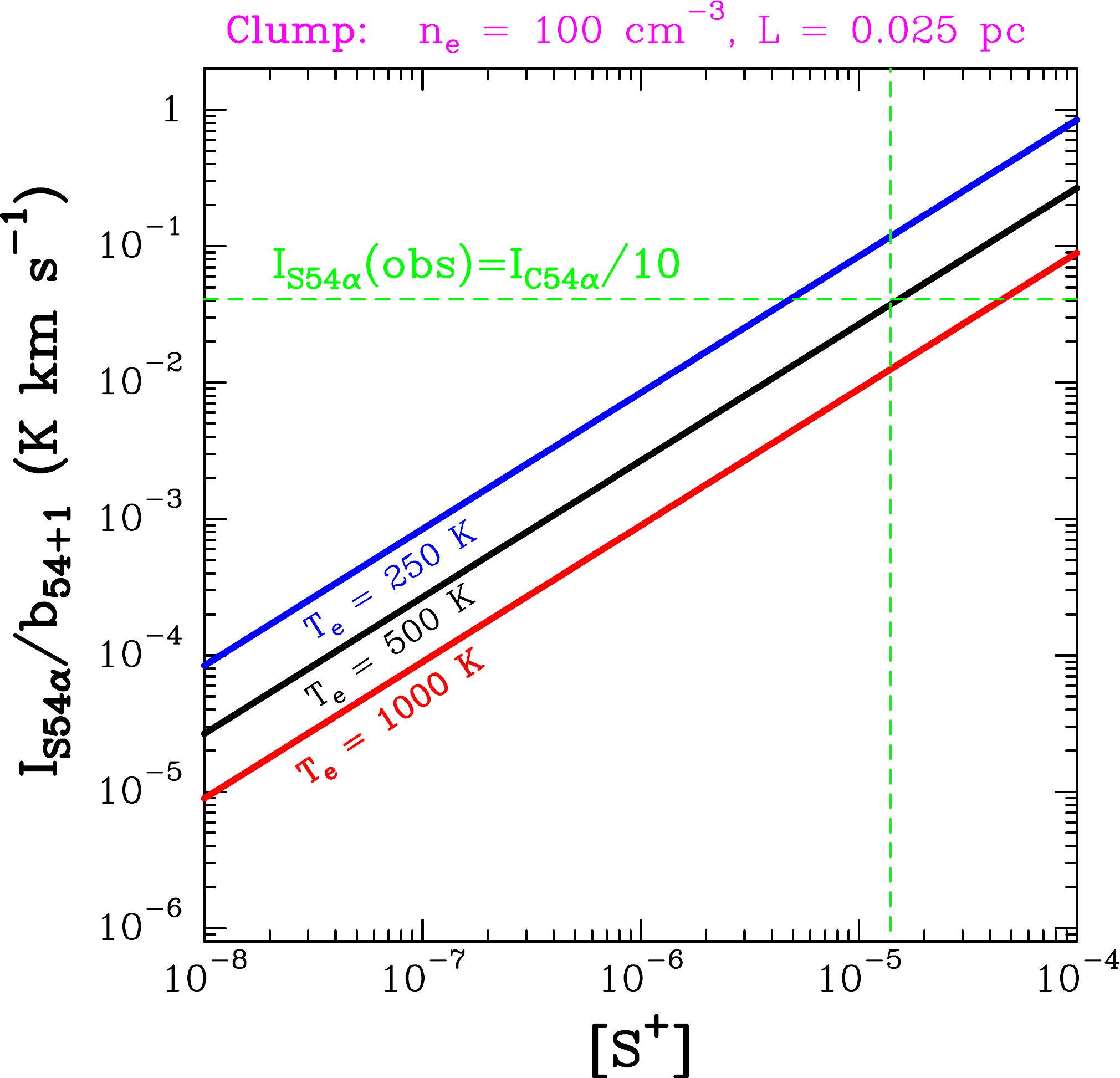}
\caption{S54$\alpha$ intensity  as a function of
sulfur abundance and $T_{\rm e}$.
\textit{Upper panel}: Interclump model of the  Bar. 
\mbox{\textit{Lower panel}}: Model that assumes that RRLs arise
from denser clumps or high-pressure gas (see text).
The  horizontal dashed green line marks the observational value.} 
\label{fig:LTE_intensities}
\end{figure}

 We  then calculated  the expected intensity of the S54$\alpha$ RRL  
using \mbox{Eq.~\ref{eq-I}} and compared it with the observational value in the Orion Bar PDR, \mbox{$I_{\rm S54\alpha}$\,$\simeq$\,$I_{\rm C54\alpha}$\,/\,$R_{\rm (54-59)\alpha}$}, where $R_{\rm (54-59)\alpha}$ is the 
\mbox{C(54-59)$\alpha$\,/\,S(54-59)$\alpha$} integrated line intensity ratio of the  stacked spectrum  \mbox{(see Table~\ref{Table-fits})}. We obtain 
\mbox{$I_{\rm S54\alpha}$\,(observations)\,=\,0.407\,/\,10.4\,=\,0.04\,K\,km\,s$^{-1}$}.
\mbox{Figure~\ref{fig:LTE_intensities}} shows a grid of single-slab models
appropriate for the two possible sources of RRL emission
discussed in the literature of the Bar: the \mbox{``interclump''} medium model -- with  \mbox{$n_{\rm e}$\,$\simeq$\,10\,cm$^{-3}$} and \mbox{$L$\,$\simeq$\,0.2\,pc} -- and the denser  ``clump'' or ``high-pressure'' model -- with  \mbox{$n_{\rm e}$\,$\simeq$\,100\,cm$^{-3}$} and \mbox{$L$\,$\simeq$\,0.025\,pc}.
Assuming a beam filling factor of one, both models reproduce the observed $I_{\rm S54\alpha}$ intensities  with a similar column
\mbox{$N$(S$^+$)\,$\simeq$\,7$\cdot$10$^{17}$\,cm$^{-2}$}
but different (LTE) electron temperatures: \mbox{$T_{\rm e}$(interclump)\,$\simeq$\,125\,K} and
\mbox{$T_{\rm e}$(clump)\,$\simeq$\,500\,K}. \citet{Cuadrado19} recently reported the detection of \mbox{3\,mm wave} \mbox{carbon RRLs} in the Orion Bar: \mbox{$\alpha$ lines from $n$\,=\,42 to 38}, \mbox{$\beta$ ($\Delta$$n$\,=\,2)} lines, and \mbox{$\gamma$ ($\Delta$$n$\,=\,3)} lines. Although they estimated, from single-slab models,  that these lines
arise from PDR layers with \mbox{$n_{\rm e}$\,$\simeq$\,60-100\,cm$^{-3}$}
and \mbox{$T_{\rm e}$\,$\simeq$\,500-600\,K}, the clump versus interclump dichotomy is not yet fully solved \citep[][]{Tielens93,Goico16}. Since PDRs have steep temperature and density gradients, we suspect that only  higher angular resolution observations will clarify the exact small spatial-scale origin of these lines.

%-------------------------------------------------------------
%-------------------------------------------------------------

The modeled column density of S$^+$, beam-averaged over the mean
angular resolution of the Yebes\,40\,m observations ($\sim$45$''$), is a factor of ten lower than the  $N$(C$^+$) independently inferred
from maps of the  [$^{13}$\CII]\,158\,$\upmu$m line  emission toward the same
angular area of the Bar: \mbox{$N$(C$^+$)\,$\simeq$\,7$\cdot$10$^{18}$\,cm$^{-2}$}  
 \citep[see Table~2 of][]{Ossenkopf13}. Therefore, the observed    \mbox{S(54-59)$\alpha$} line intensities   are consistent with both \mbox{$N$(C$^+$)\,/$N$(S$^+$)\,$\simeq$\,10}  and with \mbox{[S$^+$]\,$\simeq$\,1.4$\cdot$10$^{-5}$}. More refined estimations
 of $T_e$ and $n_e$ will require a proper non-LTE model, higher angular resolution observations, and higher S/N detections of several  RRLs.

\section{Carbon RRL observational parameters and sulfur RRL frequencies}

%-------------------------------------------------------------
\begin{figure*}[t]
\centering   
\includegraphics[scale=0.9, angle=0]{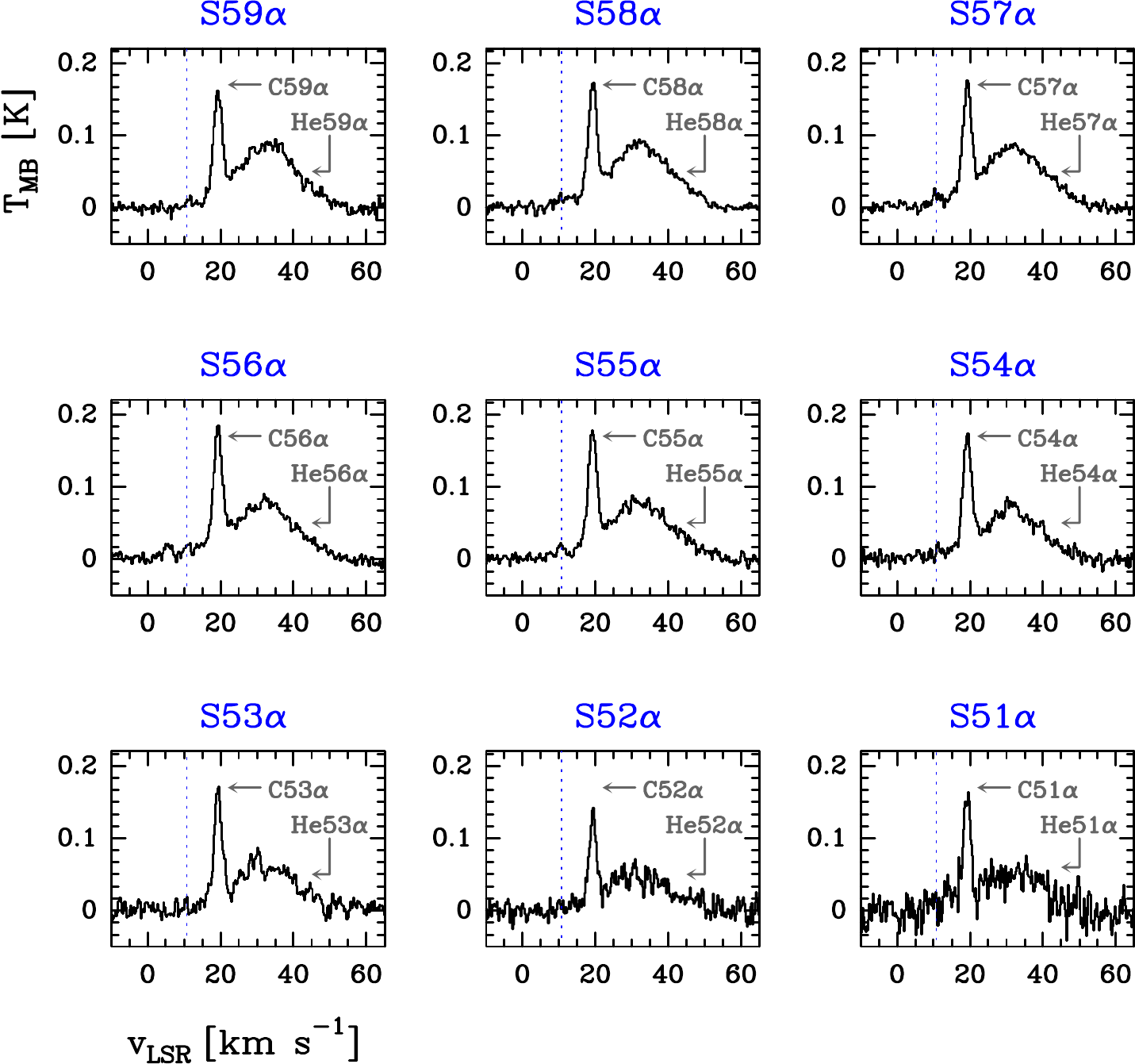}
\caption{Sulfur, carbon, and helium RRLs observed  with the Yebes\,40\,m radio telescope between 31\,GHz and 48\,GHz (at 38\,kHz resolution) toward the Orion Bar. Dashed lines indicate the position of  S$n$$\alpha$ RRLs
at $v$$_{\rm LSR}$\,$\simeq$\,10.7\,km\,s$^{-1}$, the emission of the neutral PDR gas.}
\label{fig:all_RRLs}
\end{figure*}
%-------------------------------------------------------------

%----------------------------------------------------------------------
\begin{table*}[!h] 
\vspace{2cm}
\begin{center}
\caption{C$n$$\alpha$ line spectroscopic parameters obtained  from a two-Gaussian line fit to C and He RRLs. The last column provides the frequencies
of the S$n$$\alpha$ RRLs.}  \label{Table_CRRL_Yebes}  
%\resizebox{14cm}{!}{
\begin{tabular}{c c c c c c c | c c@{\vrule height 10pt depth 5pt width 0pt}}     

\hline\hline Line     &   Frequency  &   $I=\displaystyle{\int} T_{\rm MB}$d$v^{\,a}$\,\,\,  &  $v_{\rm LSR}$    & $\Delta v$   &  $T_{\rm MB,\,Peak}^{\,a}$ & S/N$^{\,b}$  \rule[-0.3cm]{0cm}{0.8cm}\ & Line &  Frequency\\
& [GHz] & [K km s$^{-1}$]  & [km s$^{-1}$] & [km s$^{-1}$]  & [K] &  & & [GHz] \\
\hline
 C59$\alpha$   & 31.23889  &   0.410\,(0.011)  &    10.7\,(0.1)  & 2.8\,(0.1)   & 0.139 &  25 & S59$\alpha$ & 31.23978\\   
 C58$\alpha$   & 32.86859  &   0.404\,(0.009)  &    10.7\,(0.1)  & 2.6\,(0.1)   & 0.146 &  32 & S58$\alpha$ & 32.86953\\
 C57$\alpha$   & 34.61364  &   0.378\,(0.009)  &    10.7\,(0.1)  & 2.5\,(0.1)   & 0.141 &  29 & S57$\alpha$ & 34.61463\\
 C56$\alpha$   & 36.48445  &   0.399\,(0.011)  &    10.7\,(0.1)  & 2.6\,(0.1)   & 0.147 &  30 & S56$\alpha$ & 36.48550\\  
 C55$\alpha$   & 38.49255  &   0.402\,(0.010)  &    10.7\,(0.1)  & 2.6\,(0.1)   & 0.146 &  26 & S55$\alpha$ & 38.49365\\ 
 C54$\alpha$   & 40.65077  &   0.407\,(0.013)  &    10.7\,(0.1)  & 2.6\,(0.1)   & 0.146 &  22 & S54$\alpha$ & 40.65193\\   
 C53$\alpha$   & 42.97340  &   0.377\,(0.022)  &    10.7\,(0.1)  & 2.5\,(0.1)   & 0.145 &  18 & S53$\alpha$ & 42.97463\\
 C52$\alpha$   & 45.47640  &   0.260\,(0.012)  &    10.7\,(0.1)  & 2.2\,(0.1)   & 0.112 &  12 & S52$\alpha$ & 45.47770\\
 C51$\alpha$   & 48.17762  &   0.319\,(0.023)  &    10.6\,(0.1)  & 2.2\,(0.2)   & 0.138 & \,8 & S51$\alpha$ & 48.17900 \\\hline  
\end{tabular} %}                                                                                          
\end{center} 
\tablefoot{$^a$ Intensities in main beam temperature. $^b$ S/N determined from the line emission peak at a spectral resolution of 38\,kHz.
Parentheses indicate the uncertainty obtained by the Gaussian fitting routine.}                  
\end{table*}      
%----------------------------------------------------------------------

\section{Line parameters of the stacked spectrum and comparison with other lines}

%-------------------------------------------------------------
\begin{figure}[t]
\centering   
\includegraphics[scale=0.49, angle=0]{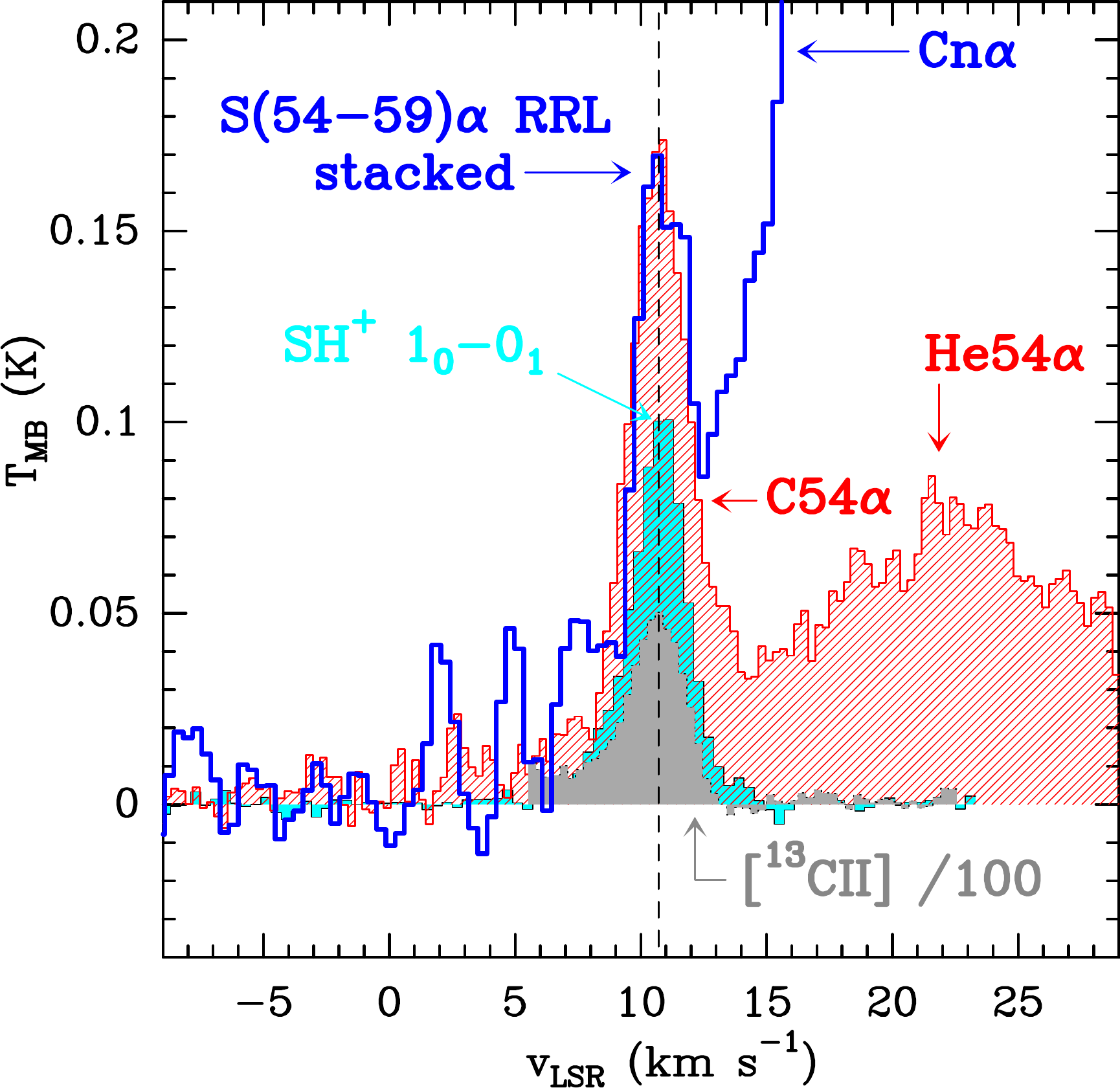}
\caption{Similar line profiles, velocity centroids, and line widths of the stacked S$n$$\alpha$ line (blue), C$54$$\alpha$ (hatched red), [$^{13}$\CII]\,158\,$\upmu$m 
(\mbox{$F$\,=\,2-1}; solid gray), and SH$^+$ (hatched cyan) lines toward the DF position (all observed or extracted at \mbox{$\theta_{\rm HPBW}$\,$\simeq$\,45$''$}  and \mbox{$\simeq$\,0.3\,km\,s$^{-1}$} resolutions). 
The  dashed  line corresponds to the LSR velocity of the PDR,  \mbox{$v_{\rm LSR}$\,=\,10.7\,km\,s$^{-1}$}.} 
\label{fig:profiles}
\end{figure}
%-------------------------------------------------------------

%----------------------------------------------------------------------
\begin{table*}[h]  
\vspace{2cm}
\begin{center}
\caption{Line spectroscopic parameters obtained from Gaussian fits toward the Orion Bar PDR.}  \label{Table-fits}  
%\begin{tabular}{l c r c c c r c c c@{\vrule height 10pt depth 5pt width 0pt}}
\begin{tabular}{l c r c c c c c    @{\vrule height 10pt depth 5pt width 0pt}}     
\hline\hline       

 Species &  Line             & Frequency       &  $\theta_{\rm HPBW}$  &  $I=\displaystyle{\int} T_{\rm MB}$\,d$v$  &  $v_{\rm LSR}$   &  $\Delta$$v$ \,\,\,\,\,\,\,     &  $T_{\rm MB,\,Peak}$  \\ 
         &                           & [GHz]           &  [$''$]                    &     [K km s$^{-1}$]  & [km s$^{-1}$]    & [km s$^{-1}$] & [K] \rule[-0.3cm]{0cm}{0.2cm}\  \\  
 \hline
  C      & (54-59)$\alpha$                       & stacked         &                                 &  2.084 (0.012)      &  10.6 (0.1)      &   2.6 (0.1)      &  0.753   \\
  S      & (54-59)$\alpha$               & stacked         &                             &  0.200 (0.011)       &  10.8 (0.1)      &   2.6 (0.2)     &  0.072 \\
\hline
SH$^+$   &  1$_0$-0$_1$ $F$=1/2-3/2  & 345.94438       &    45$^a$                   & 0.274 (0.003)        &  10.6 (0.1)      &  2.6 (0.1)      & 0.097   \\
$^{13}$C$^+$ &  $^2P_{\rm3/2}$-$^3P_{\rm1/2}$\,$F$=2-1 &  1900.46610  & 45$^a$  & 12.888 (0.812)       &  10.6 (0.1)      &  2.6 (0.2)     & 4.707  \\
\hline
  C      & 54$\alpha$                            & 40.65077        &    45                     & 0.407 (0.013)       &  10.7 (0.1)        &  2.6 (0.1)     & 0.146  \\
  He     & 54$\alpha$                            & 40.64706        &    45                     & 1.408 (0.021)       &  $-$4.7 (0.3)    &  19.5 (0.5)    & 0.069  \\ % FALTA REHACER
\hline
\end{tabular}                                                                                            
\end{center} 
\tablefoot{$^a$Extracted from [$^{13}$CII] \citep{Goico15} and SH$^+$  \citep{Goico21} maps smoothed to  $\theta_{\rm HPBW}$\,=\,45$''$ resolution (shown in Fig.~\ref{fig:map}).
Parentheses indicate the uncertainty obtained by the Gaussian fitting routine.}                  
\end{table*}      
%----------------------------------------------------------------------

\end{appendix}

\end{document}